\documentclass[12pt]{article}
\usepackage{epsfig}
\usepackage{amsmath,amssymb}

\newcommand{\be}{\begin{equation}}
\newcommand{\ee}{\end{equation}}
\newcommand{\ba}{\begin{eqnarray}}
\newcommand{\ea}{\end{eqnarray}}
\newcommand{\bea}{\begin{array}}
\newcommand{\eea}{\end{array}}

\def\tr{{\rm tr}}

\begin{document}
\begin{titlepage}
\vfill
\begin{flushright}
{\normalsize KIAS-P05035}\\
{\normalsize hep-th/0506256}\\
\end{flushright}
\vskip 1cm

\begin{center}{\LARGE New BPS Objects in {\cal N=2} Supersymmetric Gauge Theories}
\end{center}
\vskip 12mm

\begin{center}{\it \large Kimyeong Lee {\rm and} Ho-Ung Yee
\vskip 0.2in \small School of Physics, Korea Institute
for Advanced Study\\
Cheongriangri-Dong, Dongdaemun-Gu, Seoul 130-012, Korea}
\end{center}
\begin{center}
{\tt e-mail: \tt klee@kias.re.kr}, {\tt ho-ung.yee@kias.re.kr}
\end{center}


\vskip 12mm

\vskip 1cm
\begin{abstract}
We explore BPS soliton configurations in $N=2$ supersymmetric
Yang-Mills theory with matter fields arising from parallel $D3$
branes on D7 branes. Especially we focus on two parameter family
of 1/8 BPS equations,  dyonic objects, and 1/8 BPS objects and
raise a possibility of absence of BPS vortices when the number of
$D3$ branes is larger than that of $D7$ branes.
\end{abstract}

\end{titlepage}

\baselineskip 18pt

\section{Introduction and Conclusion}

Recently there has been a considerable interest in BPS solitons in
Higgs phase of supersymmetric Yang-Mills theories with eight
supercharges\cite{Tong:2002hi,Isozumi:2003rp,Isozumi:2004jc,Eto:2004rz,Sakai:2005sp,Eto:2005cp}.
Almost all known BPS objects, like magnetic flux
vortices\cite{Abrikosov:1956sx,Vachaspati:1991dz,Kneipp:2001tp},
magnetic monopoles\cite{'tHooft:1974qc}, domain
walls\cite{Cvetic:1991vp,Chibisov:1997rc}, and
instantons\cite{Eto:2004rz,Belavin:1975fg,Lambert:1999ua,Lee:2002xh},
have appeared here, sometimes with a bit of twist. These theories
can allow many degenerate vacua which can be interpolated by
domain walls. With broken $U(1)$ gauge theories, one can have
magnetic flux vortex. One of the most interesting features has
been that there can be magnetic monopoles which appear as beads on
vortex strings\cite{Hindmarsh:1985xc}.

These BPS objects can be interpreted in a simple manner from
D-brane point of view\cite{Strominger:1995ac}. Especially a simple
but rich picture appears with $N$ parallel D3 branes and $N_f$ D7
branes. In this setting, one can have $N=2$ supersymmetric $U(N)$
gauge theory with $N_f$ matter hypermultiplets in the fundamental
representation and a single adjoint hypermultiplet. One can add
Fayet-Iliopoulos(FI)-terms and mass term for the matter
hypermultiplet without breaking the supersymmetry. There are
considerable work done along this line to represent the
configurations in brane picture\cite{Lee:1997vp}.

In this work, we focus on BPS equations, dyonic 1/4 BPS, and 1/8
BPS solutions. In addition, we explore BPS vortex equations when
$N=2, N_f=1$ and found the cases where   there are no vortex
solutions of unit or double vorticity.

By studying the known bosonic BPS equations, we found that there
are two parameter family  of 1/8 BPS equations in 3+1 dimension
modulo spatial rotation and $SU(2)_R\times U(1)_R$. The FI term
breaks $SU(2)_R$ to $U(1)$ and the mass terms for matter
hypermultiplet breaks $U(1)_R$ completely. One would expect more
general BPS configurations in this setting.

Dyonic objects mean objects carrying `electric' charge. Of course
there will be no isolated electric charge in the Higgs phase due
to screening.  Electrically charged solitons could be interpreted
as  composites of soliton with fundamental strings whose ends
carry electric charge. In Higgs phase the electric charge is
neutralized by electric charge carried by the Higgs field. As the
Higgs fields carry global  flavor charge, the conserved flavor
charge instead of the total electric flux would appear in the BPS
energy formula. Besides dyonic monopoles, we show that dyonic
domain walls  as well as dyonic composites of domain
wall-monopole-vortex are also possible. When parallel D7 branes
are not lying on a single line in their transverse space, dyonic
BPS configurations which make web-like structures are also
possible. These dyonic solutions could be interpreted as the
excitations in phase moduli of BPS objects and they belong to 1/4
BPS states.

We also look for BPS solutions preserving 1/8 of eight
supersymmetries. By exploring a small perturbation of a
homogeneous 1/4 BPS configuration in 3+1 dimensional theories, we
argue that there may be no 1/8 BPS configurations satisfying the
BPS equations. However we find easily 1/8 BPS configurations in a
theory with product gauge group $U(1)\times U(1)$ with
bi-fundamental and fundamental matter field. In this analysis, the
recently discoverd\cite{Sakai:2005sp} bound states of monopoles
and domain walls play some role.

The key aspect here is that the FI parameters breaks the $SU(2)_R$
symmetry of eight supersymmetric 5+1 dimensional theory. For a
single $U(1)$ gauge group, one can use the broken R-symmetry to
choose a single direction in $SU(2)_R$ space. However with product
gauge groups,  the FI-parameters cannot be rotated to a single
direction in general. This is what allows the presence of 1/8 BPS
configurations possible.

As there are multi BPS vortex string configuration in $U(1)$
theory with $N_f=1$, we may expect there are BPS vortex string
configuration in $U(2)$ theory with $N_f=1$. While there exist
degenerate supersymmtric vacua, we will show that classically
there exists no BPS vortex configuration with unit and double
magnetic flux.  We argue that this may imply that there exists no
BPS vortex solitons  of finite magnetic flux in the theory.

One interesting direction to explore further is the interaction
between domain walls and monopoles. (See also a recent work by
Sakai and Tong***.) In string picture, parallel D1 and D3 branes
are attracted to each other. This is not apparent from the energy
argument of a BPS monopole-vortex-domain composition. The moduli
space of domain wall-monopole separation should be analyzed
carefully to resolve the question.

Another direction is to study the moduli space dynamics of
magnetic monopoles and domain walls when some of nonabelian gauge
symmetry is restored. It would be interesting to see whether there
exists a similar restoration of symmetry in the moduli space
dynamics..

Finally, all BPS solutions we study here have extended structures
with infinite energy. There may be finite action BPS solitons in
the theory. Especially it may be possible to have finite energy
(dyonic) instantons in $R^3\times S^1$ (noncommutative) space,
which do not have diverging gauge flux\cite{Lee:2002xh}.

The plan of this paper is as follows. In Sec.2, we describe 5+1
dimensional supersymmetric Yang-Mills theories and find
supersymmetric Lagrangian and its vacuum structure. In Sec.3, we
find two parametered BPS equations, especially 1/8 BPS equations.
In Sec.4, we study dyonic solutions. In Sec.5, we study 1/8 BPS
configurations and find BPS configurations with product gauge
group. In Sec.6, we show that there exists no BPS vortex solitons
of unit and double magnetic flux when $N=2$ and $N_f=1$.

{\it Note added :} In the early stage of the draft of our paper,
we came to know that the authors of Ref.\cite{Nitta}  have worked on
the classification of 1/8 BPS equations of the similar model we
considered.

\section{Six Dimensional Case}

The vector multiplet of super Yang-Mills theory of $U(N)$ gauge
group with eight supersymmetries in six dimensions is made of
$A_M, \lambda_i\, (i=1,2), {\bf D}^a $, which are  hermitian
$N\times N$ matrix valued fields. The gaugino field $\lambda_i,
i=1,2$ is made of two eight component spinors satisfying both
chirality and symplectic Majorana conditions
\be \Gamma^6\lambda_i =\lambda_i \; (i=1,2), \;\; \lambda_i =
(i\sigma^2)_{ij} B(\lambda_j^\dagger)^T \ee
where $B$ is a matrix such that $B\Gamma^M B^{-1}=(\Gamma^M)^*$.
Due to this constraint, there are only four physical degrees of
freedom in gaugino spinor. Our choice of six dimensional Gamma
matrices are
\ba && \Gamma^0 = 1_2\otimes i\sigma^3\otimes \sigma^1, \;
\Gamma^a = \sigma^a\otimes \sigma^1\otimes \sigma^1 \; (a=1,2,3) \nonumber \\
&& \Gamma^4=1_2\otimes \sigma^2\otimes \sigma^1, \;\; \Gamma^5=
1_2\otimes 1_2\otimes \sigma^2 \ea
In addition, $\Gamma^6=\Gamma^0\Gamma^1\cdots\Gamma^5=1_2\otimes
1_2\otimes \sigma^3$. With the above choice,
\be B =-i\sigma^2\otimes 1_2\otimes \sigma^3 . \ee
%

The Lagrangian for the gauge multiplet is
\be {\cal L}_1=\tr\left( -\frac{1}{4} F_{MN}F^{MN}
-\frac{i}{2}\bar{\lambda}_i \Gamma^M D_M \lambda_i + \frac{1}{2}
({\bf D}^a)^2 \right) \ee
The supersymmetric transformation becomes
\ba && \delta A_M=i\bar{\lambda}_i \Gamma_M\epsilon_i \\
&& \delta \lambda_i = \frac{1}{2} F_{MN}\Gamma^{MN}\epsilon_i + i
{\bf D}^a
\sigma^a_{ij}\epsilon_j \\
&& \delta {\bf D}^a = \bar{\epsilon}_i \sigma^a_{ij} \Gamma^I
D_I\lambda_j \ea
where the supersymmetric parameter  $\epsilon_i$ is also a chiral
spinor and satisfies the symplectic Majorana condition. The
Lagrangian and supersymmetric transformation are compatible with
the symplectic Majorana condition. The above Lagrangian is
invariant under $SU(2)_R$ transformation, under which $\lambda_i$
and ${\bf D}^a$ belong to the fundamental and adjoint
representations, respectively.

The Lagrangian for an adjoint hypermultiplet $y_i \, (i=1,2),
\chi$ where the matter spinor is anti-chiral $\Gamma^6\chi=
-\chi$, is
\be {\cal L}_2 = \tr\left( -\frac{1}{2}D_M\bar{y}_i D^My_i +
\frac{1}{2} {\bf D}^a\sigma^a_{ij}[\bar{y}_j,y_i]
-i\bar{\chi}\Gamma^MD_M\chi +\bar{\lambda}_i[\bar{y}_i,\chi]
-\bar{\chi}[y_i,\lambda_i] \right) \ee
where $D_M y_i = \partial_M y_i -i[A_M,y_i]$. Here $y_i\, (i=1,2)$
is a doublet under $SU(2)_R$ and $\chi$ is a singlet. The
supersymmetric transformation is
\be \delta \bar{y}_i = 2i\bar{\chi}\epsilon_i , \;\; \delta\chi =
D_M y_i \Gamma^M \epsilon_i \ee

The matter hypermultiplets $q_{fi}, \psi_f$ with flavor index
$f=1,...,N_f$ belong to the fundamental representation $\bar{N}$
of the gauge group $U(N)$. As in the adjoin hypermultiplet, the
matter spinor field is anti-chiral. The Lagrangian for the matter
multiplet is
\be {\cal L}_3 = \tr\left( -\frac{1}{2}D_M\bar{q}_{fi} D^M q_{fi}
+ \frac{1}{2} {\bf D}^a\sigma^a_{ij}\bar{q}_{fj} q_{fi}
-i\bar{\psi}_f\Gamma^M D_M \psi_f +
\bar{\lambda}_i\bar{q}_{fi}\psi_{f} -\bar{\psi}_fq_{fi}\lambda_i
\right) \ee
where $D_M q_{fi}= \partial_M q_{fi} +i q_{fi} A_M$. The
supersymmetric transformation is
\be \delta \bar{q}_{fi} = 2i\bar{\psi}_f\epsilon_i , \;\;
\delta\psi_f = D_M q_{fi} \Gamma^M \epsilon_i \ee

The above Lagrangians are invariant under the $SU(2)_R$ symmetry.
For a theory with abelian gauge group, one can add the Fayet-
Iliopoulos term
\be {\cal L}_{FI} = \frac{1}{2} \tr (\zeta^a {\bf D}^a ) .\ee
If the gauge group is a product group, there would be FI-terms for
each independent $U(1)$ theory. The FI parameters $\zeta^a$ breaks
the $SU(2)_R$ symmetry explicitly  and so one can use $SU(2)_R$
symmetry to rotate them to be
\be \zeta^1=0,\; \zeta^2=0,\; \zeta^3 = v^2 \ee
with $v\ge 0$. We will use both $\zeta^a$ and parameter $v$. The
${\bf D}^a$ field is not dynamical and its field equation leads to
\be {\bf D}^a = \frac{e^2}{2}\biggl\{ \zeta^a -
\sigma^a_{ij}\biggl( [\bar{y}_j,y_i] +\bar{q}_{fj}q_{fi}\biggr)
\biggr\} \ee

The dimensional reduction to 3+1 dimension induces additional
$U(1)_R$ symmetry which is a rotation under two reduced space. The
dimensional reduction with  Scherk-Schwartz mechanism induces two
mass parameters $m_f, m'_f$ for  each flavor matter multiplet
along the reduced space. If $x^4, x^5$ is reduced, then
\be D_4 q_{fi}=iq_{fi}(A_4-m_f), \; D_5 q_{fi} =iq_{fi}(A_5-m'_f).
\ee

This theory with $U(N)$ gauge group has a simple D-brane
interpretation. It is a Yang-Mills theory on $N$ parallel D3
branes near $N_f$ D7 branes whose transverse location at $x^4,x^5$
is given by the mass parameter. The location of D3 branes along
$x^4,x^5$ direcion id given by the vacuum expectation value of
adjoint scalars $A_4,A_5$. The location of D3 branes along
transverse 4 directions in D7 branes would be decided by the
expectation value of $y_i$. The dimensional reduction to $4+1$
dimension is a bit simpler with only one mass parameter and no
additional R-symmetry. The D-brane interpretation could be D4-D8
system.

One of the vacuum condition ${\bf D}^a=0$ is the ADHM condition of
$N$ instantons on $U(N_f)$ gauge theory of noncommutative four
space. The scalar fields are denoting the separation and size of
instantons. As D3 branes act as instantons on $D7$ branes, one can
see that the vacuum moduli space modulo gauge transformation is
the moduli space of instantons when the mass parameters are turned
off. With the mass parameters turned on, every D3 brane should lie
on some D7 brane at the ground state. Thus, every eigenvalue pair
of expectation value of $(A_4,A_5)$, which is diagonal at the
vacuum, should coincide with $(m_f,m'_f)$ for some $f$.

One of the simplest vacua appears when  $N=N_f$ and  all the
eigenvalue pair  of $A_4,A_5$ are distinct, such that there is
only one D3 brane for each D7 brane. It is the so-called
color-flavor locking phase, where the matter field will have a
Higgs condensation $\langle q_{f1}\rangle_{vacuum}=v$ and the
gauge symmetry plus the flavor symmetry is spontaneously broken
down to unbroken $U(1)^N$ global symmetry.

When $N=2,N_f=1$, the vacuum moduli space would be that of two
$U(1)$ instantons on noncommutative four space\cite{Lee:2000hp},
which is the so-called Eguchi-Hanson space. In this case $y_i$
does have intrinsic nonabelian components and  the gauge group
$U(2)$ is spontaneously broken to global $U(1)$ symmetry.

\section{BPS Equations}

Classically a BPS field configuration is a bosonic field
configuration which leaves some of the supersymmetry invariant. We
consider now the supersymmetric transformation to obtain the BPS
equations.  Inspired by the bosonic BPS equations, we rewrite the
supersymmetric transformation of the gaugino field  as
\ba \delta \lambda_i &=&  \Gamma^{12} \biggl( (F_{12}-F_{34}
\Gamma^{1234})\epsilon_i - i{\bf D}^3 \Gamma^{12}\sigma^3_{ij}
\epsilon_j \biggr)
 + \Gamma^{23}\biggl((F_{23}-F_{14}\Gamma^{1234})\epsilon_i  \biggr. \nonumber \\
& &  \biggl. - i{\bf D}^1\Gamma^{23}\sigma^1_{ij}\epsilon_{j}
\biggr) + \Gamma^{31}\biggl((F_{31}-F_{24}\Gamma^{1234})\epsilon_i
  -i{\bf D}^2\Gamma^{31}\sigma^2_{ij}\epsilon_j \biggr) \nonumber \\
& & + \Gamma^{\mu 0}(F_{\mu 0} - F_{\mu 5}\Gamma^{05}) \epsilon_i
+ F_{05} \Gamma^{05} \epsilon_i \ea
As $\Gamma^4\epsilon_i = -\Gamma^{123}\Gamma^{05}\epsilon_i$, the
adjoint spinor transformation is written as
\be \delta \chi =-\Gamma^{123}\bigg(  D_1 y_i \Gamma^{23}+ D_2 y_i
\Gamma^{31} + D_3y_i\Gamma^{12}+ D_4 y_i \Gamma^{05}\biggr)
\epsilon_i + \Gamma^0(D_0y_i -D_5y_i\Gamma^{05})\epsilon_i \ee
The spinor in fundamental hypermultiplet transforms as
\ba \delta \psi_{f} &=& -\Gamma^{123}\biggl( D_1 q_{fi}\Gamma^{23}
+ D_2 q_{fi}\Gamma^{31}+ D_3 q_{fi}\Gamma^{12}+ D_4
q_{fi}\Gamma^{05}\biggr) \epsilon_i \nonumber \\
& & \;\;\;\;\;\; + \Gamma^0( D_0 q_{fi} -D_5
q_{fi}\Gamma^{05})\epsilon_i \quad.\ea

We want find some supersymmetric parameter $\epsilon_i$ such that
$\delta \lambda_i, \delta \chi, \delta \psi_f$ remain zero. On
eight independent parameters of  spinor $\epsilon_i$, we impose
three independent conditions (In the case of $N=2$ NLSM, see
\cite{Naganuma:2001pu}.),
\be \Gamma^{05}\epsilon_i=\eta \epsilon_i , \;\;
\Gamma^{12}\sigma^3_{ij}\epsilon_j= i \alpha \epsilon_i,\;\;
\Gamma^{23}\sigma^1_{ij}\epsilon_j = i\beta \epsilon_i\quad,\;\;
 \label{spinorcond}
 \ee
with $\alpha, \beta, \eta$ take $\pm 1$ independently.  Since
$\Gamma^0\Gamma^1...\Gamma^5=1$ for chiral $\epsilon_i$, these
conditions imply that
\be \Gamma^{31}\sigma^2_{ij}\epsilon_i= -i \alpha\beta \epsilon_i,
\;\; \Gamma^{1234}\epsilon_i= \eta \epsilon_i\quad. \ee
These are conditions on eight independent Majorana  parameters in
the  spinor $\epsilon_i$, as they are compatible with the
symplectic Majorana condition. If we impose any one of the
conditions, the number of independent SUSY parameters would be
reduced by one half to four of the original value. If we impose
any two of them, the number of independent SUSY parameters are
reduced to two or 1/4 of the original one. If we impose all three
of them, the number of independent parameters is reduced to one,
1/8 of the original value.

One can obtain different conditions by six dimensional Lorentz
transformations and $SU(2)_R$ transformations. In reduction to
$3+1$ dimensions, only nontrivial ones modulo remaining symmetries
is the rotation between the remaining coordinates and the reduced
coordinates. In the reduction to $3+1 $ dimensions of coordinate
$x^0,x^1,x^2,x^3$, the above condition can be generalized to new
spinor conditions with two parameters,
\ba && \Gamma^0(\Gamma^5\cos\theta +\Gamma^3\sin\theta)\epsilon_i
=
\eta \epsilon_i, \;\; \Gamma^1(\Gamma^2\cos\varphi +
 \Gamma^4\sin\varphi)\sigma^3_{ij} \epsilon_{ij}=  i\alpha \epsilon_i \,,\nonumber \\
&& (\Gamma^2\cos\varphi+\Gamma^4\sin\varphi)(\Gamma^3\cos\theta
-\Gamma^5 \sin\theta)\sigma^1_{ij}\epsilon_j = i \beta\epsilon_i
\label{spinorcond2}\ea
This implies that
\be (\Gamma^3\cos\theta-\Gamma^5\sin\theta)\Gamma^1
\sigma^2_{ij}\epsilon_j=-i\alpha\beta\epsilon_i, \;\;
\Gamma^{124}(-\Gamma^3\cos\theta
+\Gamma^5\sin\theta)\epsilon_i=\eta \epsilon_i \,, \ee
Note also $D_4q_{fi}=iq_{fi}(A_4-m_f)$ and
$D_5q_{fi}=iq_{fi}(A_5-m'_f)$. In reduction to $4+1$, we can put
$\varphi=0$ as it is a part of four dimensional spatial rotation.

We use the generalized spinor condition (\ref{spinorcond2}) to
find the BPS equations satisfied by the bosonic configurations for
the minimum amount 1/8 of the original supersymmetries. For any
vector with spatial indices, we introduce barred indices so that
\ba && V_{\bar{1}}=V_1, \; V_{\bar{2}}=
V_2\cos\varphi+V_4\sin\varphi,\;
V_{\bar{3}} = V_3\cos\theta -V_5\sin\theta, \nonumber \\
&& V_{\bar{4}}= V_4\cos\varphi-V_2\sin\varphi, \;
V_{\bar{5}}=V_5\cos\theta+ V_3\sin\theta \ea
 From $\delta \lambda_i=0$, we get the gauge field part of BPS
equations,
\ba && F_{0\bar{5}}=0 , \; F_{\bar{\mu}0}-\eta F_{\bar{\mu}
\bar{5}}=0 \; (\mu=1,...4),\; F_{1\bar{2}} - \eta
F_{\bar{3}\bar{4}} + \alpha {\bf D}^3=0,\;  \nonumber
\\
&& F_{\bar{2}\bar{3}}-\eta F_{1\bar{4}}+\beta {\bf D}^1 = 0, \;
F_{\bar{3}1}-\eta F_{\bar{2}\bar{4}}-\alpha\beta {\bf D}^2=0 , \;
\label{BPSeq1}\ea
From $\delta\chi=0$ and $\delta \psi_f=0$, we also obtain
\ba && \beta D_1 y_j\sigma^1_{ji}-\alpha\beta D_{\bar{2}} y_j
\sigma^2_{ji}+ \alpha D_{\bar{3}} y_j
\sigma^3_{ji} -i\eta D_{\bar{4}} y_i  =0,  \nonumber \\
&&  D_0y_i -\eta D_{\bar{5}} y_i= 0 , \;\;\;
D_0q_{fi}-\eta D_{\bar{5}} q_{fi} = 0 , \nonumber\\
&& \beta D_1q_{fj}\sigma^1_{ji} -\alpha\beta
D_{\bar{2}}q_{fj}\sigma^2_{ji} +\alpha
D_{\bar{3}}q_{fj}\sigma^3_{ji}- i\eta D_{\bar{4}}q_{fi} =0
\label{BPSeq2} \ea
These are the BPS equations for 1/8 BPS configurations. The BPS
equations preserving more supersymmetry  can be obtained by
imposing additional conditions to the above BPS equations. For
example, 1/4 BPS configurations satisfy two sets of 1/8 BPS
equations with, say, both $\alpha=1$ and $\alpha=-1$. There is
also a Gauss law constraint for the BPS configurations,
\be -\frac{1}{e^2} \sum_{\mu=0}^5 D_{\bar{\mu}}  F_{\bar{\mu} 0}
-\frac{i}{2}([\bar{y}_i,D_0y_i]-[D_0\bar{y}_i,y_i])
-\frac{i}{2}(\bar{q}_{fi}D_0q_{fi}-D_0\bar{q}_{fi}q_{fi})= 0
\label{Gaussl}\,. \ee

Using the BPS equation, the central charge\cite{Witten:1978mh} for
the BPS energy bound can be found to be
\ba Z\!\! &=&\!\! \frac{1}{2}\int d^3x\; \tr \biggl(
\frac{\eta}{2e^2} F_{\bar{\mu}\bar{\nu}}
\tilde{F}_{\bar{\mu}\bar{\nu}} -\alpha \zeta^3( F_{1\bar{2}}-\eta
F_{\bar{3}\bar{4}}) -\beta \zeta^1( F_{\bar{2}\bar{3}}
-\eta F_{1\bar{4}})  \biggr. \nonumber \\
& & \;\;\;  \;\;  \biggl. + \alpha\beta \zeta^2 (F_{\bar{3}1}-\eta
F_{\bar{2}\bar{4}}) \biggr) +\eta m_f Q_f \cos\theta + \eta T_{03}
\sin\theta  + Z' \label{central}\ea
where $\mu,\nu=1,2,3,4$ and
$\tilde{F}_{\mu\nu}=\frac{1}{2}\epsilon_{\mu\nu\rho\sigma}F_{\rho\sigma}
$.  After dimensional reduction to (3+1) dimensions,
$D_4y_i=-i[A_4,y_i]$ and $D_4 q_{fi}=i q_{fi} (A_4-m_f)$, and so
$F_{14}= D_1 A_4$ and $F_{45}= -i[A_4,A_5]$. The charge $Q_f$ is
the one carried by the $f$'th-flavor matter field,
\be Q_f = \frac{i}{2} \int d^3x \; \tr
(\bar{q}_{fi}D_0q_{fi}-D_0\bar{q}_{fi}q_{fi})\quad, \ee
and $T_{03}$ is the linear momentum along $x^3$ direction,
\be T_{03}= \frac{1}{2}\int d^3x\; \tr  \biggl( \frac{1}{e^2}
\sum_{\mu=1,2,4,5} F_{\mu 0}F_{\mu 3} +
(D_0\bar{y}_iD_3y_i+D_3\bar{y}_iD_0y_i)+ (D_0\bar{q}_{fi}D_3
q_{fi} +D_3\bar{q}_{fi}D_0q_{fi}) \biggr)\,. \ee
The boundary term $Z'$ is given by
\be Z' = \int d^3x \biggl( \eta \partial_i \tr (F_{i0}A_5)
\cos\theta  +\cdots \biggr)\quad, \ee
where $\dots$ indicates the terms quadratic in matter fields and
expected to have zero boundary contribution in both Coulomb and
Higgs phases. The first part would have nontrivial contribution in
the Coulomb phase where there would be nontrivial electric field.)

The above BPS equations and the energy bound are complicated
functions of two parameters $\varphi$ and $\theta$. For example, a
complication arises as
\be
F_{\bar{3}\bar{4}}=F_{34}\cos\theta\cos\varphi-F_{32}\cos\theta\sin\varphi
-F_{54}\sin\theta\cos\varphi+F_{52}\sin\theta\sin\varphi \ee
Using the un-bared coordinate indices, we note that the first term
of the above expression can be expressed as
\be  F_{\bar{\mu}\bar{\nu}} \tilde{F}_{\bar{\mu}\bar{\nu}} =
F_{\mu\nu}\tilde{F}_{\mu\nu}\cos\theta +
4(F_{12}F_{45}+F_{24}F_{15}+F_{41}F_{25})\sin\theta \ee
There are also the boundary terms depending on quark fields, which
is supposed to make vanishing contributions almost all cases.

Once we fix $\zeta^a=v^2\delta_{3}^a$, which is possible for the
theories of $U(N)$ gauge group but not for those with product
gauge group like $U(1)\times U(1)$, the BPS energy does not
depends on the choice of the parameter $\beta$. This means that
1/4 BPS configurations defined by $\alpha$ and $\eta$ parameters
could have 1/8 BPS excitations without generating additional
energy, which is strange. Indeed we see that this is impossible in
some simple case studied in  Sec.5.

We can choose two parameters $\theta, \varphi$ to be arbitrary. If
we fix $\zeta^a$, we no longer have the freedom of $SU(2)_R$
transformation, and the parameters $\theta,\varphi$ become
physically meaningful. One typical cases of BPS equations would be
when $\theta=\varphi=0$. In this case, the barred spacial indices
become the un-barred ones and $\partial_4=\partial_5=0$. The other
extreme may be when $\theta=\varphi=\pi/2$. In this case the time
dependent part becomes $F_{03}=0, (D_0-\eta D_3)\;  {any \;
field}=0$, and
\ba && \eta F_{12}+i[A_4,A_5]+ \beta{\bf D}^1=0,\;\; D_1 A_4 +\eta
D_2
A_5 +\alpha {\bf D^3}=0,  \nonumber \\
&& D_1 A_5-\eta D_2 A_4 -\alpha\beta{\bf D}^2=0, \\
&&\bigl(\beta D_1 y_j \sigma^1_{ji} +i\eta D_2 y_i +i\alpha\beta
[A_4,y_j]\sigma^2_{ji} + i\alpha [A_5, y_j]\sigma^3_{ji}\bigr)=0,
\;\;\;\nonumber\\&&\bigl(\beta D_1 q_{fj} \sigma^1_{ji} +i\eta D_2
q_{fi}   -i\alpha\beta q_{fj}(A_4-m_f)\sigma^2_{ji} -i\alpha
q_{fj}(A_5-m'_f)\sigma^3_{ji} \bigr)=0 \nonumber\label{bps3}\ea

We know quite a bit of the topological objects of the theories in
$\theta=\varphi=0$. The simplest object is a 1/2 BPS vortex
soliton along $x^3$ direction in $U(1)$ theory with
$N_f=1$\cite{Kneipp:2001tp}. It satisfies the BPS equation with
$\beta=-1$,
\be  2F_{12} = v^2 - |q_1|^2,\; (D_1-iD_2)q_1=0\quad, \ee
where $y_i=0, q_2=0$, dropping the flavor index. Especially  a
unit flux vortex has a vortex tension $T_v =  \pi v^2  $. This
could be regarded as a D1 string on a single D3 brane in a single
D7 brane. The next simplest object is a 1/2 BPS domain wall
parallel to $(x^1,x^2)$
plane\cite{Tong:2002hi,Isozumi:2003rp,Isozumi:2004jc,Abraham:1992vb}.
With $N=1$ and $N_f=2$ with two different $m_f$ along $x^4$
direction, the 1/2 BPS equations with $\alpha\beta=1$ becomes
\be 2 \partial_3 A_4= v^2- \sum_{f=1}^{2}|q_{f1}|^2 \,, \quad
\partial_3 q_{f1}=-q_{f1}(A_4-m_f)\quad, \ee
where $y_i=0, q_{f2}=0$,  $m_1<m_2$, and $A_5=0$. The $A_4$
interpolates between $m_1$ and $m_2$. It describes the D3 brane on
first D7 brane interpolating to the second D7 brane. The wall
tension is $ T_{12}= \pi v^2 (m_2-m_1) $.

More complicated object is a 1/4 BPS configuration made of
magnetic monopole beads in a vortex flux
tube\cite{Hindmarsh:1985xc}. With $N=N_f=2$ and in the
color-flavor locking phase with $m_1<m_2$ and $A_5=m'_f=0$, the D1
string on the first D3-D7 branes can interpolate to the second
D3-D7 branes. The D1 string connecting two D3 branes appears as a
magnetic monopole. In the Higgs phase, the magnetic flux is
confined to flux string and so the 1/4 BPS object is made of two
vortices emerging opposite to the magnetic monopole, where two
$U(1)$'s of $U(2)$ flux are carried to opposite direction. The
composite has the energy of a simple sum of vortex tension and
monopole mass.

Most complicated 1/4 BPS object is a composite made of vortex and
domain walls, which also allows some magnetic
monopoles\cite{Sakai:2005sp,Carroll:1997pz,Auzzi:2005yw}. With
$\beta=-1, \alpha=-1$, from the BPS energy one notices that with
positive $\tr F_{12}$ and $\tr F_{34}$, which means postive vortex
flux and domain wall charge where $A_4$ is increasing, there is
negative instanton energy, or monopole energy. This is the
so-called bound energy of vortex-domain wall\cite{Sakai:2005sp}.
If a vortex terminates at the domain wall, the wall shape gets
deformed in large distance away from the contact point. The detail
has been also studied recently\cite{Auzzi:2005yw}. Of course one
can add additional monopole kink to this vortex-domain wall
junction, which carries the positive monopole energy. In some
cases, the magnetic monopole can pass the domain wall. When a
vortex penetrating a domain wall is deformed so that the contact
points at the both sides of the domain wall do not coincide to the
same point, the monopole could not pass the domain wall due to the
energy consideration, which means that there could be repulsive
potential at the domain wall. It would be interesting to find
whether our conjecture is true.

A typical  solution of the BPS equations of $\theta=\varphi=\pi/2$
would be the 1/4 BPS domain wall
junction\cite{Eto:2005cp,Abraham:1990nz,Saffin:1999au} with $N=1,
N_f=3$. Suppose that the three complex masses $m_f+im'_f$ lie  on
vertices of an equitriangle so that $m_f+im'_f = me^{2\pi i f/3}$
with $f=1,2,3$. The BPS equation would be give by (\ref{bps3})
with $\zeta^3=v^2$ and the wall junction would lie on on $x^1,x^2$
plane with $x^3$ translation invariance. The ansatiz is that
$y_i=0, q_{f2}=0, A_1=A_2=A_3=0, \partial_3=0$ and the BPS
equation becomes
\ba && \partial_1 A_4+\eta \partial_2 A_5=
-\frac{\alpha}{2}(v^2-|q_{f1}|^2), \; \partial_1 A_5-\eta \partial_2 A_4=0 \\
&& \partial_1q_{f1}-\alpha(A_4-m_f)q_{f1}=0,\; \eta \partial_2
q_{f1} -\alpha (A_5-m_f')q_{f1}=0 \ea
The web of wall solutions of this type in a bit more complicated
setting has been also studied recently\cite{Eto:2005cp}.

\section{Lorentz Boosted, or  Dyonic  Solutions}

For the BPS configurations, the time dependent part can be solved
with
\be A_0=\eta ( A_5\cos\theta + A_3\sin\theta) , \; \;  \partial_0
q_{fi} -\eta( \partial_3 \sin\theta - i m'_f\cos\theta) q_{fi} =0
\ee
while  $(\partial_0-\partial_3 \sin\theta)  =0$ for any field in
the adjoint representation.  One can see that it is a Lorentz
boost along $x^3$ axis with velocity $v= \sin\theta$ when $|\theta
|<\pi/2$. However, the $\theta=\pi/2$ case is still physically
distinct as it cannot be obtained through finite boost. The Gauss
law is also equivalently Lorentz boosted version. This matches
with the energy being increased with $T_{03} v = {\cal O}(v^2)$
for small $v$ as $T_{03}$ itself is linear in $v$ for small $v$.
For the domain wall junctions, $T_{03}=0$ with $\theta=\pi/2$ due
to the $x^3$ translation invariance of the configuration. Thus one
cannot boost them along $x^3$, but may be able to put some
massless wave along $x^3$ without breaking the supersymmetry
further.

When $\theta=0$,  $A_0=\eta A_5$ and the all adjoint fields are
time-independent and $\partial_0 q_{fi}+ i\eta m'_f q_{fi}=0$. The
$f$-th flavor charge becomes
\ba Q_f &=& \eta \int d^4x \; \tr \biggl( (m'_f-A_5)
\bar{q}_{fi}q_{fi}\biggr).  \ea
As the total electric charge vanishes in the Higgs phase, we put
the constraint $\sum_f Q_f=0$. Here we consider the fundamental
string connecting D3 branes with net $U(1)=\tr(U(N))$ charge
vanishes in the Higgs phase. The energy carried by the flavor
charge becomes
\be E_Q = \eta \sum_f m'_f Q_f = \sum_f \int d^3 x \;  m'_f \, \tr
\biggl( (m'_f-A_5) \bar{q}_{fi}q_{fi}\biggr)   \ee
For most of BPS objects considered here, they have moduli space
parameter corresponding to a global phase rotation. The excitation
along this direction would lead to the dyonic solutions. The Gauss
law would give the equation for $A_5$ which is exactly the zero
mode equation satisfied by the phase moduli coordinate in the
background gauge of solutions without $A_5$, $m'_f$ included. The
parameters $m'_f$ serve as coefficients of the excited phase
moduli direction vector of the dyonic solution.

Consider a vortex-monopole composite with $N=N_f=2$ in a
color-flavor locking phase with $m_1<m_2$. One can impose
additional BPS condition on electric charge section without
breaking any additional supersymmetry. One has to solve the above
Gauss law which can be solved in priciple in this monopole-vortex
background. The result describes a composite of D1-fundamental
strings connecting D3 branes, which means that the monopole
carries electric charge.  However, the $A_5$ would approach
exponentially vacuum expectation value away from the monopole
region, implying that the electric charge is shielded by the Higgs
field. For two flavor case, one can choose $A_5 \sim A_4$ up to
constant shift as $(m_f, m'_f)$  lies along a line. Note that $E_Q
\sim (\Delta m')^2$ and $Q_2-Q_1 \sim \Delta m'$, and so the
relative flavor charge fixes $m'_2-m'_1$ as in the dyons in
Coulomb phase.

When $N_f\ge 3$,  $D7$ branes does not need to lie on a line as
three points given by the mass parameters do not lie along a line
in general. In this case one could have a web of D1, F1 and
$(p,q)$ strings\cite{Bergman:1997yw}. For example consider
$N=N_f=3$ in the color-flavor locking phase. If the D7 branes are
separated from each other and lie on almost straight line, one can
imagine a D1 string interpolating two D3 branes at the end. When
we introduce the fundamental strings connecting, say first and
second D3 branes, the resulting configuration would be a vortex
string where there are two fundamental monopoles attacted to each
other, but the Coulomb repulsion due to the electric charge in
short distance keeps them away from each other. This is quite
similar to the corresponding configuration in the Coulomb phase.
The key difference would be that in the Higgs phase there may be
no upper bound on F1 string numbers as the electric repulsion
would be shielded in large separation.

It is straightforward to extend this to situations of multiple
domain walls. Consider $N=2$, $N_f=3$ with two domain walls
interpolating $m_1,m_2$ by first D3 and $m_2,m_3$ by second D3
($m_1<m_2<m_3$). If we turn on $m'_2$ slightly, these two domain
walls are attracted, and it is balanced by giving them electric
charges proportional to $m'_2$ distributed on their world volume.
This would be web-like structure of D3 branes and sheet of
fundamental strings, attached to D7 branes.

Another dyonic BPS configuration is possible.  Start with a 1/2
BPS domain wall of a single D3 brane, interpolating two D7 branes
in position. Fundamental strings connecting two D7 branes at the
wall generates the electric dipole on D3 brane. Two ends of the
dipole are shielded by the Higgs field of different flavor, and so
the configuration has the Higgs charge. One needs to solve the
Gauss law in the domain wall background. From the domain wall
world sheet point of view, the fundamental F1 string appears as a
charge of phase or magnetic flux on effective 2+1 dimensional
theory. Uniform charge configuration would corresponds to unform
magnetic flux configuration on effective 2+1 dimensional theory.

In our BPS equation there is additional parameter $\varphi$. To
see its role in $N=1, N_f=2$ with $\zeta^a=v^2\delta_{a3}$,
$A_5=m'_f=0$, let us consider the domain wall solution with $2, 4$
directions mixed. With only dependence on $x^1$ and $x^3$ and
$A_1=A_2=A_3=0$, $\eta=\alpha=-1$, $\theta=0$, BPS equations
(\ref{BPSeq1}) and (\ref{BPSeq2}) for $A_4$ become
\ba && (\partial_3\cos\varphi-\partial_1\sin\varphi) A_4=
\frac{1}{2} (v^2-\sum_f |q_{f1}|^2), \nonumber\\
&& (\partial_3\cos\varphi-\partial_1\sin\varphi)q_{f1} + q_{f1}
(A_4-m_f)=0\nonumber\\ &&
(\partial_3\sin\varphi+\partial_1\cos\varphi)A_4=0,\quad
(\partial_3\sin\varphi+\partial_1\cos\varphi)q_{f1}=0\quad.
 \ea
This corresponds to a spatial rotation in $(x^1,x^3)$ plane. The
origin of this fact can be traced back to the correlation between
$(x^2,x^4)$ and $(x^1,x^3)$ in the spinor projection conditions.

\section{1/8 BPS Objects in Theories with Product Gauge Groups}

While we found 1/8 BPS equations which seems to be general up to
six dimensional Lorentz boost and $SU(2)_R$ symmetry, it is not
clear whether 1/8 BPS configurations are allowed. After the
dimensional reduction to 3+1 dimensions with two general angle
parameters, one cannot make arbitrary six dimensional rotation,
especially $F_{45}=0$ in $U(1)$ theory. While we are interested in
the general characteristics of 1/8 BPS configurations, if any
exists, it seems very hard to solve the BPS equations.

Let us start with a theory with a simple gauge group, say, $U(N)$.
To find out what  the characteristics of 1/8 BPS configurations
are, let us start with BPS configuration of constant field
strength with zero matter expectation value. From BPS equations
for the gauge fields (\ref{BPSeq1}) for the constant field
strength, we can make $SU(2)_R$ rotation to put the FI parameter
to 3-th direction and $SU(2)$ spatial rotation in $x^1,
x^{\bar{2}}, x^{\bar{3}}$, which rotates both $\epsilon_i$ and the
gauge field strength $F_{\bar{\mu}\bar{\nu}} $ with
$\mu,\nu=1,2,3,4$. From this one can see that the constant field
configuration is at most 1/4 BPS configuration.

Inhomogeneous BPS field configuration can be obtained by
extracting magnetic fluxes from the system. To see whether 1/8 BPS
configurations are possible when the field configuration is
inhomogeneous in space, we ask whether 1/8 BPS perturbation arises
in 1/4 BPS homogeneous background\cite{Lee:1995ei}.

Let us start with a $U(1)$ gauge theory on $3+1$ dimension with
single flavor. Let us start with a 1/4 BPS configuration which is
homogeneous in space and time with $A_0=A_5$ and $\eta=\alpha =-1$
with $\theta=\varphi=0$. The FI term becomes ${\bf D}^a=
e^2v^2/2\;\; \delta_{a3}$ and and we choose the constant 1/4 BPS
field strenghs to be
\be F_{12}= \frac{e^2v^2}{2}\, a , \;\; F_{34}= \frac{e^2v^2}{2}\,
(1-a),\;\; F_{23}=\frac{e^2v^2}{2}\, b,\;\;
F_{14}=-\frac{e^2v^2}{2} \, b \ee
with constants $a,b$. This is a generalization of many previously
known homogenous solutions. The homogeneous BPS configuration in
$U(1)$ Higgs model with single Higgs field represents the unform
distribution of vortices on plane, which has the critical total
magnetic flux\cite{Lee:1995ei}. In $SU(2)$ gauge theory, one could
have magnetic monopole sheet or homogenous field configuration
with uniform instanton density.  The energy density is then
 \be {\cal E} =  \frac{e^2v^4}{4}\biggl(1 +b^2-a(1-a)
 \biggr) \ee
In four dimensions, the contribution from the intersection of
$F_{12}$ and $F_{34}$ can decrease the tension when $0<a<1$ and
can be regarded as an anti-selfdual instanton part with the
negative energy, which can be regarded as a bound energy of two
uniform magnetic flux. Note that the minimum energy is positive.

In 3+1 dimensions, it represents the bound energy of a domain wall
and infinite number of vortex strings penetrating domain walls.
The number of flavors does not play any role. For $b>0$ and $a$
not in this interval induces self-dual instanton density which
contributes positive energy.  Note that there are critical total
flux $e^2v^2/2$ in our unit. From the brane point of view, the
above BPS solution induces D3 branes with homogeneous field on its
world sheet, tilted with respect to D7 branes.

We want to see whether there is any 1/8 BPS deformation of this
homogeneous configuration. The BPS equation implies that there
should be nonzero $q_i, i=1,2$ for 1/8 BPS configurations, which
we regard as a small perturbation. (Here we drop flavor index as
there is only one flavor.) We solve the 1/8 BPS equation by the
perturbation expansion with $\beta =-1$. To the first order we
first solve the matter BPS equation in the unform background,
\be D_1
q_j\sigma^1_{ji}+D_2q_j\sigma^2_{ji}+D_3q_j\sigma^3_{ji}-iD_4 q_i
=0,  \ee
We choose the gauge
\be   A_1=0, \; \; A_2=\frac{e^2v^2}{2} (ax^1-bx^3), \; A_3=0 , \;
A_4=\frac{e^2v^2}{2}((1-a)x^3-bx^1) \ee
The above equation is satisfied if
\be \partial_1 q_i + \frac{e^2v^2}{2} x^1 q_j(
a\sigma^3-b\sigma^1)_{ji}=0 , \;\; \partial_3q_i +
\frac{e^2v^2}{2} x^3q_j(b\sigma^1+(1-a)\sigma^3)_{ji}=0 \ee
One can convince oneself that only $q_1$ becomes normalizable
along both $x^1$ and $x^3$ directions  for $b=0$ and $0<a<1$ while
$q_2$ is not normalizable at all. For $1/8$ BPS deviation, we need
both normalizable $q_1$ and $q_2$ modes to start the perturbative
approach and so there is no $1/8$ BPS deviation from the 1/4 BPS
configuration. The BPS equation for the gauge fields indicate the
second order effect of the $q_1$ perturbation  reducing the total
magnetic flux and instanton or monopole number. Thus one can guess
that the above homogeneous configuration, while remaining 1/4 BPS,
is continuously connected to the two intersecting flux sheet along
$x^1-x^2$ and $x^3-x^4 $ plane with finite magnetic monopole
charge and negative bound energy. In the brane picture, the end
result would be  the intersection of D3 brane domain wall and D1
string.

While the above analysis does not provide clear picture about the
existence of 1/8 BPS configurations in 8 supersymmetric $U(N)$
gauge theories, it suggests that  1/8 BPS configurations are
unlikely.

Now consider a theory with $U(1)\times U(1)$ gauge group with
fundamental matter fields in each gauge group and also  many
bi-fundamental matter fields of charge $(+1,-1)$. Let assume that
two FI parameters are not parallel and so, say,
$\zeta^{(1)a}=\delta_{a3}$ and $\zeta^{(2)a}=\delta_{a1}$. (Here
we put the proportional numbers and electric charges to be one for
simplicity.)  If there is no bi-fundamental matter fields, two
theories are not interacting and so it is obvious that there can
be 1/8 BPS configurations. They can be made of 1/4 BPS
configurations of each gauge group but they are not aligned and so
break the supersymmetry further to 1/8. Even when bi-fundamental
fields exist, such 1/8 BPS configurations are possible if
bi-fundametnal field has zero expectation value.

  To see whether bi-fundamental matter field can develope any nontrivial
expectation value, let us start with 1/8 BPS homogeneous
configuration in this theory of two product gauge group,
\ba && F^{(1)}_{12}=a,\; F^{(1)}_{34}=1-a \\
&& F_{23}^{(2)}=b,\;\; F_{14}^{(2)}=1-b \ea
The energy density of the configuration becomes
\be {\cal E}=\frac{1}{4}(2-a(1-a)-b(1-b)) \ee
With the gauge
\be A_2^{(1)}=ax^1, \; A^{(1)}=(1-a)x^3, \; A^{(2)}_2 = -bx^3 ,
\;\; A_4^{(2)}=(1-b)x^1 \ee
The interesting question is whether there exists a nonzero mode
for the bi-fundamental field $q_i$, whose BPS equation is
satisfied if
\ba && \partial_1 q_i +x^1 q_j(a\sigma^3-(1-b)\sigma^1)_{ji}=0 \\
&& \partial_3 q_i+x^3 q_j((1-a)\sigma^3-b\sigma^1)_{ji} \ea
The normalizable solution along $x^1,x^3$ direction is possible if
$a=b=1/2$, in which case two matrices are proportional to each
other and so can be exponentiated easily. Once we found this
normalizable zero mode, we can feed it to the BPS equation for the
gauge field, which leads to the second order perturbation, which
reduces the sum of the magnetic fluxes. Of course there will be
also nontrivial BPS deformation of fundamental matter field for
each gauge group. One can imagine the continuous deflation of the
total flux would lead to some sort of intersecting U(1) magnetic
vortex sheets, while remaining  1/8 BPS. From 1/4 BPS case, one
can see that first U(1) vortex line along $x^3$ direction meets a
first U(1) domain parallel to the  $1-2$ plane. The second U(1)
vortex line along $x^1$ direction meets a second U(1) domain wall
parallel to the $2-3$ plane.Together they would remain 1/8 BPS. In
addition, there would be nontrivial bi-fundamental matter field in
this 1/8 BPS configuration, making two configurations to be
connected together.

\section{Nonexistence of BPS Vortices}

Most of the analysis on solitons so far have been done when
$N_f\ge N$. Especially there would be no supersymmetric vacua if
$N_f<N$ without adjoint hypermultiplet. When $N_f< N$, the adjoint
hypermultiplet plays a crucial role for supersymmetric vacua to
exist. When $N=2, N_f=1$, the explicit  vacuum solution modulo
local gauge transformations is known\cite{Lee:2000hp}. At the
vacuum the scalars in vector multiplet $(A_4,A_5)
 = (m_1,m'_1) $, proportional to the identity matrix.
With adjoint hypermultiplet, the vacuum equation ${\bf D}^a=0$ is
the ADHM condition on two instantons in noncommutative $U(1)$
theory, and the moduli space metric becomes the Eguchi-Hanson
space. It is depending on eight parameters, four of which are the
position of the center of mass of two D3 branes in D7 branes, and
so flat and does not affect our analysis. There are additional
four parameters which indicates the relative distance and phase
between two D3 branes in D7 branes. Due to the FI term, there
would be Higgs condensation on D3 branes. Explicitly, \ba &&
\langle \bar{y}_1 \rangle = w_1+\frac{z_1}{2}\left(
\begin{array}{cc} 1 &
\sqrt{\frac{2b}{a}} \\ 0 & -1 \end{array}\right) , \;\; \langle
y_2 \rangle = w_2 + \frac{z_2}{2}\left(\begin{array}{cc} 1 &
\sqrt{\frac{2b}{a}} \\ 0 & -1 \end{array} \right) \\
&& \langle \bar{q}_1 \rangle = v \left(\begin{array}{c} \sqrt{1-b}
\\ \sqrt{1+b} \end{array} \right), \;\; \langle q_2 \rangle =0 \ea
where $a = (|z_1|^2+|z_2|^2)/(2v^2) $ and $b=\sqrt{a^2+1}-a$. The
vacuum moduli space is characterized by four complex parameteres
$w_i,z_i$. The parameter $w_i$ denotes the location of the center
of mass points of two D3 branes on D7 background and the parameter
$z_i$ denotes the relative position.

We know there are BPS multi-vortex solutions when $N=N_f=1$. The
question is whether any BPS vortex solitons exist when $N=2,
N_f=1$.  Suppose we put a single $D1$ string on one of $D3$ branes
when two D3 branes are in infinite separation. Clearly it is BPS.
As we change vacuum moduli parameters so that two D3 branes are
almost on top of each other, we may expect that there would be 1/2
BPS vortex solutions. To see whether it is true, we look at a
consistent ansatz.

Rather the surprise appears when two D3 branes are on top of each
other, or when the vacuum moduli is at minimum two sphere of
Eguch-Hanson space. In this case the consistent ansatz becomes
\be  \;\; \bar{y}_1 = \left(\begin{array}{cc} 0 & Z \\ 0 & 0
\end{array}\right), \;\; y_2=0, \;\; \bar{q}_1 = \left(\begin{array}{c} 0,\\
Q_2
\end{array}\right), \;\; q_2=0, \;\; A_1+iA_2= {\rm diag} (A,B)
\ee
The BPS equation get simplified to be (
$\partial={1\over2}(\partial_1-i\partial_2)$)
\ba && \partial Z-i(A-B) Z=0, \; \partial Q_2-i B Q_2 = 0 \,,\\
&& -i(\bar{\partial} A -\partial \bar{A}) = v^2-|Z|^2, \\
&& -i(\bar{\partial}B- \partial \bar{B}) = v^2+|Z|^2-|Q|^2 \,.\ea
Asymptotic value of $|Z|^2 $ and $|Q|^2$ are $v^2$ and $2v^2$,
respectively. The above BPS equations can be combined to
\ba && -\partial_i^2 \ln |Z/Q|^2 = v^2-|Z|^2 \\
&& -\partial_i^2 \ln |Q|^2 = v^2+|Z|^2- |Q|^2 \ea

The BPS energy is determined by $\frac{v^2}{2}(F_A+F_B )=
2v^2-|Z|^2$. The vorticity of $Z$ and $Q_2$ are $l_1, l_2$, then
the flux $\int d^2 x(F_A -F_B)= 2\pi l_1 >0$ and $\int d^2x F_B=
2\pi l_2>0$. The $\int d^2x F_A = 2\pi (l_1+l_2)$ and the energy
is $\pi v^2 (l_1+l_2)$. From examining the above equations, one
can easily draw the fact that there is no solution with $l_1=0,
l_2>0$ or $l_1>0,l_2=0$ or $l_1=l_2>0$. The only possibility is
$l_1-1\ge l_2\ge 1$.  As we move D3 branes apart, it suggests that
there is no BPS configurations possible for vortices with
vorticity 1 or 2 even D3 branes are apart. Assuming that the
continuity of the BPS configurations here as we do not see any
critical separation between D3 branes matter, there seems to be
only one logical conclusion, that is, that two D3 branes with any
parallel D1 string on them become {\it repulsive}. That means
there is no BPS configuration with any vorticity and finite
separation. This seems to be only consistent result. It would be
interesting to verify this conjecture.

\vskip 1cm \centerline{\large \bf Acknowledgement} \vskip 0.5cm

We thank Muneto Nitta for correspondence about his work, and
Norisuke Sakai, Piljin Yi for helpful discussions. 
K.L. is supported in part by KOSEF R01-2003-000-10319-0
and KOSEF through CQUeST at Sogang University.
H.U.Y is
supported by grant No. R01-2003-000-10391-0 from the Basic
Research Program of the Korea Science \& Engineering Foundation.

\end{document}